\documentclass[journal]{IEEEtran}

\ifCLASSINFOpdf

\else

\fi

\usepackage[english]{babel}

\makeatletter
\adddialect\l@ENGLISH\l@english
\makeatother

\usepackage[usenames]{color}
\usepackage{times}
\usepackage{graphicx}
\usepackage{url}
\usepackage{subfigure}
\usepackage{subfigure}
\usepackage{url}
\usepackage{color}
\usepackage{pifont}
\usepackage{amsmath,amssymb,amsfonts}
\usepackage[utf8]{inputenc}

\begin{document}

\title{PANAS-t: A Pychometric Scale for Measuring Sentiments on Twitter}

\author{Pollyanna Gon\c{c}alves$^{\star}$~~~~~~~Fabr\'icio Benevenuto$^{\star\dag}$~~~~~~~~~Meeyoung Cha$^\ddag$\\~\\
	$^{\dag}$Computer Science Department, Federal University of Ouro Preto, Brazil\\
	$^{\star}$Computer Science Department, Federal University of Minas Gerais, Brazil\\
    $^{\ddag}$Graduate School of Culture Technology, KAIST, Korea\\
}

\markboth{Systems, Man, and Cybernetics, Part C: Applications and Reviews,~Vol.~XX, No.~XX, June~2013}%
{Gonçalves \MakeLowercase{\textit{et al.}}: PANAS-t: A Pychometric Scale for Measuring Sentiments on Twitter}

\maketitle


\begin{abstract}
Online social networks have become a major communication platform, where people share their thoughts and opinions about any topic real-time. The short text updates people post in
these network contain emotions and moods, which when measured collectively can unveil the public mood at population level and have exciting implications for businesses,
governments, and societies. Therefore, there is an urgent need for developing solid methods for accurately measuring moods from large-scale social media data. In this paper, we
propose PANAS-t, which  measures sentiments from short text updates in Twitter based on a well-established psychometric scale, PANAS (Positive and Negative Affect Schedule). We
test the efficacy of PANAS-t over 10 real notable events drawn from 1.8 billion tweets and demonstrate that it can efficiently capture the expected sentiments of a wide variety of
issues spanning tragedies, technology releases, political debates, and healthcare.

\end{abstract}

\begin{IEEEkeywords}
Twitter, sentiment analysis, public emotion, public mood, psychometric scales, PANAS.
\end{IEEEkeywords}

\section{Introduction}

\noindent
Online social networks (OSNs) like Facebook and Twitter have become an important communication platform, where people share their thoughts and opinions about any topic in a collaborative manner in real-time. As of 2012, Facebook has over one billion active users, which is one-seventh of the world population, and Twitter similarly has over 400 million registered users each producing hundreds of millions of status updates every day ~\cite{twitter}. Given the scale and the richness of these networks, the potential for mining the data within OSNs and utilizing observations from such data is tremendous, and OSN data have been a gold mine for scholars in fields like linguistics, sociology, and psychology who are looking for real-time language data to analyze~\cite{Lazer2009}.

The massive-scale detailed human lifelog data found in OSNs have important implications for businesses, governments, and societies. The following areas of research demonstrate directly how useful observations from mining OSN data could be. First, social media data can be used to find resonance of important real-time debates and breaking news. As more people are seamlessly connected to the Web and OSN sites by mobile devices, people participate in delivering and propagating prominent and urgent information like political uprising~\cite{iran.twitter,Cheong}, natural disasters~\cite{Sakaki@www10}, and the upheaval of epidemics~\cite{gomide2010dengue}. Second, social media data can be used to   not only understand the current trends but also predict future trends such as movie sales ~\cite{DBLP:journals/corr/abs-1003-5699}, political elections~\cite{Tumasjan,
Diakopoulos:2010:CDP:1753326.1753504,citeulike:7044833}, as well as stock market~\cite{Bollen}.

The key features of OSN data that allow for the above implications is at their immediacy and immensity, for which the development of new methods on large-scale and real-time collection and analysis of OSN data are crucial. One such important development is at inferring sentiments in OSNs. A recent work has showed that real-time moods of people can be gauged on a global level, instead of relying on questionnaires and other laborious and time-consuming methods of data collection~\cite{Golder.science.2011}. Measuring sentiments from unstructured OSN data can not only broaden our understanding of the human nature, but also comprehend how, when, and why individuals' feelings fluctuate according to various social and economic events.

While sentiment analysis in OSNs is getting great attention, existing work on measuring sentiments from OSN data has focused on extracting opinions (not feelings) for marketing purposes~\cite{Pang+Lee:08b} and on finding correlation of moods with some other factor such as happiness~\cite{Dodds-Happiness} and stock price~\cite{Bollen}. Most research on inferring moods from social media texts have directly employed existing natural language processing tools like LIWC (Linguistic Inquiry and Word Count)~\cite{Pennebaker.liwc.2003}, PANAS (Positive and Negative Affect Schedule)~\cite{Watson,Watson-X},  ANEW (Affective Norms for English Words)~\cite{Kim}, and Profile of Mood States (POMS)~\cite{Bollen} that have been developed to suit more traditional style writing, such as formal articles that uses proper language (but not for unstructured and less-formal OSN data). However, relatively little attention has been paid developing solid methods for adjusting existing natural language processing tools for specific types of OSN data.

In this paper,  we use well-established psychometric scales, PANAS, to measure sentiments from short text updates in Twitter and propose PANAS-t, which is an eleven-sentiment psychometric scale adapted to the context of Twitter. PANAS-t contains positive and negative mood states and is suitable to measure sentiments about any sort of event in Twitter. To establish PANAS-t, we used empirical data from a unique dataset containing 1.8 billion tweets. We used such data to compute normalization scores for each sentiment, so that any increase or decrease in positive or negative moods over time can be measured relatively to the presence of the overall sentiments in this dataset. This approach makes PANAS-t very simple and practical to be used for large amounts of data and even for real-time analysis.

To validate our approach, we extracted 10 real notable events that span a wide variety of issues spanning tragedies, technology releases, political debates, and healthcare from the 3.5 years worth of Twitter data, and demonstrated that PANAS-t can effectively capture the mood fluctuations during these events. The 10 events studied include the 2009 Presidential election in the US, death of the singer Michael Jackson, as well as the natural disasters like the 2010 Earthquake in Haiti. Our qualitative evaluation offers strong evidences that PANAS-t correctly captured expected sentiments for the analyzed events.

The remainder of this paper is organized as follows. Section 2 surveys existing approaches to measure sentiments from text. Section 3 details how PANAS-t works and Section 4 describes the Twitter dataset. Section 5 provides experimental evidences that our approach is able to capture public mood from tweets associated to noteworthy events. Finally, Section 6 concludes the paper and offers directions for future work.

\section{Related Work}

\noindent
With the growth of social networking on web, sentiment analysis and opinion mining have become a subject of study for many researches. In this section, we survey different techniques used to measure sentiments from online text and describe related work that studied sentiments in Twitter.

Several methodologies have being used by researchers to extract sentiment from online text.  An overview of a number of these approaches was well-presented in Pang and Lee's
survey~\cite{Pang+Lee:08b}, which covers several methods that use Natural Language Processing (NLP) techniques for sentiment analysis---techniques by which subjective properties of text are
inferred using statistical methods.  Those methods are usually suitable for constructing sentiment-aware and opinion mining Web applications, which analyze feedback of
consumers or users about a particular product or service~\cite{Aue+Gamon:05a,Airoldi+Bai+Padman:06a}.

Chesley \textit{et al.}~\cite{Chesley+al:06a} utilized verbs and adjectives extracted from Wikipedia to classify text from blogs into three categories: objective, subjective-positive,
or subjective-negative. The verb classes used in the paper can express objectivity and polarity (i.e., a positive or negative opinion), and the polarity of adjectives can be drawn from their entries in the online dictionary, with high accuracy rates of two verb classes demonstrating polarity near 90\%. More recently, Pak and Paroubek~\cite{Pak} utilized strategies of grammatical structures' recognition to define if a tweet written by a user is a subjective phrase or not. They demonstrated that superlative adjectives, verbs in first person, and personal pronouns are often used for expressing emotions and opinions as opposed to comparative adjectives, common, and proper nouns that are a strong indicator of an objective text.

Other approaches that extract sentiment from online text rely on machine learning, a technique in which algorithms learn a classification model from a set of previously labeled data, and then apply the
acquired knowledge to classify text new into sentiment  categories. In~\cite{Bermingham:2010:CSM:1871437.1871741}, the authors use Support Vector Machine (SVM) and Multinomial Naive Bayes (MNB) classifiers to test whether brevity in microblog posts give any advantage in classifying sentiment and in fact find that short document length suggests a more compact and explicit sentiment than long document length. In~\cite{pcalaisKDD11}, the authors use Random Walk (RW)-based model and compare it with SVM to predict bias in user opinions. Although these approaches are applicable for several scenarios, supervised learning techniques require manual intervention for pre-classifying training data, which may be  infeasible for massive-scale social media data.

Another line of research on extracting sentiments from online text is at measuring a happiness index from text~\cite{Golder.science.2011}. Dodds and Danforth~\cite{Dodds-Happiness} proposed a method that computes the level of happiness of an unstructured text. They showed that while the happiness index inferred from song lyrics trends downward from the 1960s to the mid 1990s remained stable within genres, that of blogs has steadily increased from 2005 to 2009. While providing new insights, one drawback of this approach is that the happiness index proposed has a single scale and do not provide any other categorization of rich sentiments, which is the focus of this work.

Miyoshi, T.~\cite{miyoshi:sentiment} \textit{et al}. propose a method to estimate the semantic orientation of Japanese reviews about some target products. Authors selected words that possible change the semantic orientation of a text and then concluded if the review of a product can be considered desirable or not. In order to evaluat their approach, authors analyzed 1,400 Japanese reviews of eletric products such as LCD and MP3 Players in order to separated it in positive and negative reviews.

There are two studies that are more closely related to our goals. Kim \textit{et al.}~\cite{Kim} proposed a method for detecting emotions using Affective Norms for English Words (ANEW), which is a dataset that contains normative emotional ratings for 1034 English words. Each word in the ANEW dataset is associated with a rating of 1--9 along each of three dimensions: valence, arousal, and dominance. Based on these scales, the authors examined sample tweets about celebrity deaths and found ANEW to be a promising tool mine Twitter data. Another study~\cite{Bollen} utilized {Profile of Mood States (POMS)}, which is a psychological rating scale that measures certain mood states consisting of 65 adjectives that qualify 6 negative feelings: tension, depression, anger, vigor, fatigue and confusion. The authors applied this scale to identify sentiments on a sample of tweets and evaluate the mood of users related to market fluctuations and events like political elections in the United States.

This paper builds upon the above efforts and adopt a different psychometric scale called PANAS (Positive and Negative Affect Schedule)~\cite{Watson,Watson-X} to achieve new contributions. First, compared to the machine learning-based or other dictionary-based approaches, PANAS contains a well-balanced set of both positive and negative affects. This makes PANAS suitable to analyze reactions of people not only on crisis events such as celebrity deaths and natural disasters, but also amusing events that incur positive emotions. Second, compared to existing work that tested sentiment extraction on sample data, we use the complete data gathered from Twitter to test the idea, which allows us to perform appropriate normalization to adjust PANAS for Twitter.

\section{PANAS-t: Affect Measure for Twitter}

\noindent
Our approach to measure sentiments in Twitter is rooted on a well-known psychometric scale, namely PANAS. We begin by describing PANAS-x, a popular expanded version of PANAS, which we utilize and then  describe the normalization steps that we take to adapt the psychometric scale for Twitter.

\subsection{The PANAS and PANAS-x Scales}

\noindent
The original PANAS consists of two 10-item mood scales and was developed by Watson and Clark~\cite{Watson} to provide brief measures of PA (Positive Affect) and NA (Negative Affect). Respondents are asked to rate the extent to which they have experienced each particular emotion within a specified time period (typically during the past week), with reference to a 5-point scale. Ever since the development of the test, the words appearing in the checklist broadly tapped the affective lexicon. Later, the same authors developed an expanded version by including 60 items. The expanded version, called PANAS-x, not only measures the two original higher order scales (PA and NA), but also 11 specific affects: Fear, Sadness, Guilt, Hostility, Shyness, Fatigue, Surprise, Joviality, Self-Assurance, Attentiveness, and Serenity.

Table~\ref{tab:table2} summarizes the word composition of the PANAS-x scale~\cite{Watson-X}. The negative affect includes words like ``afraid,'' ``scared,'' and ``nervous,''  while the fatigue affect state includes words like ``sleepy,'' ``tired,'' and ``sluggish.'' The items in PANAS-x has been validated extensively and also is known to have strongly relationship with POMS categories, with convergent correlations ranging above 0.85.  In addition,  PANAS-x has been demonstrated with its excellence over POMS, because the items in PANAS-x tend to be less highly correlated with one another, and thus show better discriminant validity. For instance, the mean correlation among the PANAS-x Fear, Hostility, Sadness, and Fatigue
scales was 0.45, which is significantly lower than the mean correlation (0.60) among the corresponding POMS scales.

The authors also validated that individual trait scores on the PANAS-X scales (a) are stable over time, (b) show significant convergent and discriminant validity when
correlated with peer-judgments, (c) are highly correlated with corresponding measures of aggregated state affect, and (d) are strongly and systematically related to measures
of personality and emotionality~\cite{Watson-X}. Due to this excellence, we choose to adopt PANAS-x for analyzing short text updates from online social media.

\begin{table*}[t]
			\centering
			\small
				\begin{tabular}{| l | l |  p{5cm} | }
				\hline
				\textbf{General Dimension Scales} & \\
				Negative Affect (10) & afraid, scared, nervous, jittery, irritable, hostile, guilty, ashamed, upset, distressed. \\
				Positive Affet (10) & active, alert, attentive, determined, enthusiastic, excited, inspired, interested, pround, strong. \\ 
				&\\
				\textbf{Basic Negative Emotions Scales} & \\
				Fear (6) & afraid, scared, frightened, nervous, jittery, shaky. \\
				Hostility (6) & angry, hostile, irritable, scornful, disgusted, loathing.\\
				Guilt (6) & guilty, ashamed, blameworthy, angry at self, disgusted with self, dissatisfied with self.\\
				Sadness (5) & sad, blue, downhearted, alone, lonely.\\ 
				&\\
				\textbf{Basic Positive Emotions Scales} & \\
				Joviality (8) & happy, joyful, delighted, cheerful, excited, enthusiastic, lively, energetic.\\
				Self-assurance (6) & proud, strong, confident, bold, daring, fearless.\\
				Attentiveness (4) & alert, attentiveness, concentrating, determined.\\ 
				&\\
				\textbf{Other Affective States} & \\
				Shyness (4) & shy, bashful, sheepish, timid.\\
				Fatigue (4) & sleepy, tired, sluggish, drowsy.\\
				Serenity (3) & calm, relaxed, at ease.\\
				Surprise (3) & amazed, surprised, astonished.\\
				&\\
				& \textit{Note}. The number of terms comprising each scale is shown in parentheses.\\
				\hline
				\end{tabular}
				\caption{Item composition of the \textit{PANAS-x} scales.}
				\label{tab:table2}
\end{table*}

\subsection{Adjusting PANAS-x for Twitter}

\noindent
Tweets expressing certain sentiments may appear more frequently than others, leading to a bias or dominance of a small set of sentiments in OSN data. Thus,  in order to tell if tweets expressing a specific type of sentiment has increased or decreased for a given event (e.g., celebrity death or natural disasters), we first need to know what kinds of sentiments appear during ``typical'' or non-event periods. Unfortunately, it is hard or impossible to determine which dates would be classified as such. One natural baseline would be to aggregate sentiments over a long period of time and consider the proportion of each type of sentiment as the baseline. Therefore, by comparing the proportion of tweets that
contain a specific sentiment during a given event against the entire baseline, one can know how sentiments have changed \textit{related} to the presence of a given event in the
entire dataset.

We describe the methods to compute the baselines for comparison. We assume each normalized tweet can be mapped to a single sentiment. When a tweet contains any of the adjectives in Table~\ref{tab:table2}, we associate the corresponding sentiment $s$ as the main sentiment of the tweet. In case none of the sentiment words in Table~\ref{tab:table2} appear in a tweet, we cannot infer the sentiment for that tweet. This limitation is common to most other sentiment tools described in the related work. In case there is a tie and more than two sentiments can be found in a single tweet, we choose the first sentiment that appears in the tweet (based on the locatio of the adjectives) as the major sentiment of that tweet, although such ties are very rare and hence are negligible for analysis.

The baseline sentiment can be then calculated as follows. Let $T$ be the entire set of normalized tweets and $T_s$ the subset of these tweets related to sentiment $s$. The baseline value for each sentiment, $\alpha_s$, is defined as the proportion that divides the number of occurrences of tweets of each type of sentiment by the total number of normalized tweets in our dataset:
\begin{equation}
\alpha_s = \frac{|T_s|}{|T|}
\end{equation}

Table~\ref{tab:table3} shows the baseline values for all 11 sentiments in PANAS-x from the 3.5 years worth of Twitter data, which we will describe in detail in the next section. Some sentiments occur orders of magnitude more frequently than others. Tweets expressing \textit{fatigue} occurs nearly 32 more frequently than tweets expressing \textit{shyness}. This skew in frequency indicates that normalization is needed to comprehend the effective change of a given sentiment, because treating the any increase in the number of fatigue and shyness tweets equally will result in under-estimation and over-estimation of these sentiments, respectively. Therefore, the inherent skew in sentiments reinforces that a proper normalization specific to the OSN is necessary.

\begin{table}[h]
			\centering
				\begin{tabular}{ | l | r |}
				\hline
				\textbf{Sentiment ($s$)} & Baseline (\textbf{ $\alpha_s$})\\
				\hline
				\hline
				
				Fear & 0.0063791\\
				\hline
				Sadness & 0.0086279\\
				\hline
				Guilt & 0.0021756\\
				\hline
				Hostility & 0.0018225\\
				\hline
				Shyness & 0.0007608\\
				\hline
				Fatigue & 0.0240757\\
				\hline
				Surprise & 0.0084612\\
				\hline
				Joviality & 0.0182421\\
				\hline
				Self-assurance & 0.0036012\\
				\hline
				Attentiveness & 0.0008997\\
				\hline
				Serenity & 0.0022914\\
				\hline
				\end{tabular}
				\caption{Fraction of tweets for each sentiment in the entire dataset.}
				\label{tab:table3}
\end{table}

Given the baseline sentiment values in Table~\ref{tab:table3}, we can now compute the relative increase or decrease in sentiments for a particular sample of tweets as follows.  Let $S$ be the set of tweets (e.g., natural disaster) and $S_s$ the subset of these tweets related to sentiment $s$. We define $\beta_s$ as the relative occurrence of sentiment $s$ for the event $S$ and compute it as follows:
\begin{equation}
\beta_s = \frac{|S_s|}{|S|}
\end{equation}
Finally, we define the PANAS-t score as an eleven-dimensional sentiment vector, where the PANAS-t score function $P(s)$ for sentiment $s$ is computed as bellow:
\begin{equation}
 P(s) =
  \begin{cases}
   \frac{(\alpha_s - \beta_s)}{\alpha_s} 						&\text{if } \beta_s \le {\alpha_s} \\
   -\frac{(\beta_s - \alpha_s)}{\beta_s}				  &\text{otherwise}
  \end{cases}
\end{equation}

The value of  $P(s)$ varies between -1 and 1 for each sentiment $s$. An event with $P(fear)$ = 0 means that the event has no increase or decrease for the sentiment
\textit{fear} in comparison with the entire dataset of tweets posted as of 2009.  A positive value of 0.3 would mean an increase of 30\%, and so on. Our strategy to compute the PANAS-t score is simple and suitable for allowing the comparison of both the increase and decrease for each type of sentiment relatively to a non-bias dataset. More importantly, Table~\ref{tab:table3} provides a baseline for comparison against any kinds of sample tweets.  For instance, one could easily crawl tweet samples using the Twitter API
and normalize the sentiment scores found with our baselines.

\subsection{Most popular words of PANAS-t}

\noindent
Having seen that the level of baseline sentiments in tweets are skewed, we quantify which words of the PANAS-t scales appear most frequently in the dataset. Table~\ref{tab:table7} shows the frequency of each adjective based on the entire Twitter data. Even within a given sentiment, certain adjectives are used more frequently to express feelings. The most popular adjectives are ``sleepy'' in the fatigue category (appearing over 8.0 million times), followed by ``happy" in the joviality category (appearing over 3.8 million times). Other popular words include ``tired'', ``excited'', ``sad'', ``amazed'', ``alone'', and ``surprised'', which all appear more than 1 million times.

However, certain words in the PANAS-x scales are rarely used in Twitter to express the moods, such as ``downherted'' in the sadness category and ``blameworth'' in the guilt category. We may expect that not all words in the PANAS-x will appear frequently in OSNs, because the PANAS-x scale was originally designed to be used in a different environment (i.e., intrusive surveys). A patient submitted to PANAS test needs to mark in a scale from 1 to 5 how much each of these words tell about her mood state. Despite of this difference between PANAS-x and PANAS-t, the next section presents a number of situations in which PANAS-t can capture the expected mood states of populations about a number of noteworthy events accurately.

\begin{table*}[t]
	\centering
    \begin{tabular}{|l|l|l|}
        \hline
        \textbf{Self-assurance} & \textbf{Attentiveness}  & \textbf{Fatigue}  \\ \hline
      	proud: 762,990 & alert: 209,062 & sleepy: 8,043,591 \\
        strong: 596,376 & concentrating: 123,725 & tired: 3,486,574 \\
        daring: 295,047 & determined: 96,616 & sluggish: 19,938\\
        confident: 95,858 & attentive: 5,456 & drowsy: 18,435\\
        bold: 90,101 & & \\
        fearless: 20,084 & & \\

        \hline \hline
        \textbf{Guilt} & \textbf{Fear} & \textbf{Sadness} \\ \hline
        ashamed: 492,371 & scare: 1,649,193 & sad: 2,765,458\\
        guilty: 324,446 & nervous: 668,867 & alone: 1,096,592\\
        angry at self: 7,873 & afraid: 515,224 & lonely: 15,858\\
        disgusted with self: 2,853 & shaky: 173,142 & blue: 987\\
        dissatisfied with self: 61 & frightened: 75,260 & downhearted: 286\\
        blameworthy: 19 & jittery: 12,791 &\\

        \hline \hline
       	\textbf{Hostility} & \textbf{Joviality} & \textbf{Serenity}\\ \hline
       	angry: 483,937 & happy: 3,802,662 & at ease: 1,030,236  \\
       	irritable: 268,546 & excited: 3,170,837 & relaxed: 737,668  \\
        disgusted: 220,470 & delighted: 117,074 & calm: 258,576  \\
       	loathing: 72,330 & lively: 43,552 &  \\
       	hostile: 12,614 & enthusiastic: 34,323 & \\
       	scornful: 7,516 & energic: 22,159  & \\
       	& joyful: 21,663  & \\
       	& cheerful: 19,178 & \\
       	
       	\hline \hline
       	\textbf{Surprise} & \textbf{Shyness} & -\\ \hline
       	amazed: 2,758,114 & shy: 320,611 & \\
       	surprised: 1,050,164 & timid: 13,521 & \\
       	astonished: 19,047 & bashful: 2,556 & \\
       	& sheepish: 6,850 & \\
       	\hline
    \end{tabular}
    \caption{Frequency of each term of PANAS-t in the total database.}
	\label{tab:table7}
\end{table*}

\section{Twitter dataset}

\noindent
The dataset used in this work includes extensive data from a previous measurement study that included a complete snapshot of the Twitter social network and the complete history of
tweets posted by all users as of August 2009~\cite{cha_icwsm10}.  More specifically, the dataset contains {54,981,152}  users who had 1,963,263,821 follow links among
themselves and posted  {1,755,925,520} tweets (as of August 2009).  Out of all users, nearly 8\% of the accounts were set as private, which implies that only their friends
could view their links and tweets. We ignore these users in our analysis.

This dataset is appropriate for the purpose of this work for the following reasons.  First, the dataset contains all users with accounts created before August 2009. Thus, it is not
based on sampling techniques that can introduce bias towards some characteristics of the users.  Second, this dataset contains all tweets of these users, which is essential for measuring the increase or decrease of a certain sentiment related to tweets of a specific event. Thus, this dataset uniquely allows us to normalize the presence of
sentiments of a sample of tweets relatively to the inherit sentiments in Twitter.

\subsection{Data cleaning steps}

\noindent
In order to analyze only those tweets that possibly express individuals' feelings, we only into account tweets that contain explicit statements of their author’s mood states by matching the following expressions in tweets: \textit{``I'm"}, \textit{``I am"}, \textit{``I"}, \textit{``am"}, \textit{``feeling"}, \textit{``me"} and \textit{``myself"}. A similar approach has been used in~\cite{Bollen} in finding correlations of Twitter moods and stock price. In total, we found  {479,356,536} tweets that match these patterns, which correspond to about 27\%
of the entire dataset of tweets.

Once we found a set of candidate tweets that contain emotions and moods, we further cleaned the data as follows. We first applied common language processing approaches such as case-folding, stemming, and
removal of stop words, URLs, and common verb-forms. We then separated  individual terms using white-space as delimiters and also removed  commas, dashes, and others non-alphanumeric characters.  For example, a tweet \textit{``I am so scared about swine flu"} terns into the following set of terms, [\textit{I, am, scare, swine, flu}]. In the remainder of this paper, we use the above described normalization and analyze a total of {479,356,536} normalized tweets.

\section{Evaluation of PANAS-t}

\noindent
In order to evaluate the extent to which PANAS-t can accurately measure sentiments of Twitter users, we need ground truth data to compare the results with our methods.  Such ground truth data is difficult to
obtain because sentiments are subjective by nature. In this paper, we consider a few number of strategies to perform this evaluation. First we evaluate a set of popular events, for which
the sentiments associated with them are expected or easy to be verified. Second, we compare our results obtained using PANAS-t with an analysis performed using common emoticons
most used by users for express their feeling on social networkings. Third, we show that the baseline values computed for PANAS-t were useful to measure sentiments from a dataset of
tweets collected in a different period.

\begin{table*}[t]
		\centering
		\begin{tabular}{| p{1,9cm} | p{1,7cm} | p{6,0cm} | r |}
		\hline
		\textbf{Event} & \textbf{Duration} & \textbf{Description (Example keywords)} & \textbf{\# Tweets}\\
		\hline
		\hline
				
		\textsc{H1N1} & Mar 1 -- Jul 31, 2009 & Disease outbreak (tamiflu, outbreak, influenza, pandemia, pandemic, h1n1, swine, world health organization) & 335,969  \\ \hline
		\textsc{AirFrance} & Jun 1--6, 2009 & A plane crash (victims, passengers, A330, 447, crash, airplane, airfrance) & 29,765  \\ \hline
		\textsc{US-Elec} & Nov 2--6, 2008 & US presidential election (clinton, biden, palin, vote, mccain, democrat, republican, obama) & 185,477  \\ \hline
		\textsc{Obama} & Jan 18--22, 2009 & Presidential inauguration speech (barack obama, white house, presidential, inauguration) & 43,015  \\ \hline
		\textsc{Michael-Jackson} & Jun 25--30, 2009 & Death of celebrity (rip, mj, michael jackson, death, king of pop, overdose, drugs, heart attack, conrad murray) & 56,259  \\ \hline
		\textsc{Susan-Boyle} & Apr 11--16, 2009 & Appearance of a new celebrity (susan boyle, I dreamed a dream, britain's got talent) & 7,142  \\ \hline
		\textsc{Harry-Potter} & Jul 13--17, 2009 & Release of a movie (harry potter, half-blood prince, rowling) & 194,356  \\ \hline
		\textsc{Olympics} & Aug 6--26, 2008 & Beijing Olympics (olympics, medals, china, beijing, sports, peking, sponsor) & 12,815   \\ \hline
		\textsc{Samoa} & Sep 28 -- Oct 4, 2009 & Natural disaster (tsunami, samoa islands, tonga, earthquake) & 23,881   \\ \hline
		\textsc{Haiti} & Jan 11--17, 2010 & Natural disaster (haiti, earthquake, richter, port au prince, jacmel, leogane) & 236,096 \\ \hline
				
	\end{tabular}
	\caption{Summary of events that were analyzed.}
	\label{tab:table4}
\end{table*}

\subsection{Testing across popular real-world events}

\noindent
We picked nine events that were widely reported to have been covered by Twitter\footnote{Top Twitter trends \url{http://tinyurl.com/yb4965e}}. These events, summarized in
Table~\ref{tab:table4}, span topics related to tragedies, products and movie releases, politics, health, as well as sport events. To extract tweets relevant to the these events, we
first identified a set of keywords describing each topic by consulting news websites, blogs, wikipedia, and informed individuals. Given the selected list of keywords, we
identified the topics by searching for keywords in the tweet dataset.  We limited the duration of each event because popular keywords are typically hijacked by spammers after
certain time~\cite{benevenuto@ceas10,benevenuto@ceas11}. Table~\ref{tab:table4} also displays the keywords used and the total number of tweets for each topic.

\begin{figure*}[t]
		\centering
			\subfigure[H1N1]{\includegraphics[width=0.32\textwidth]{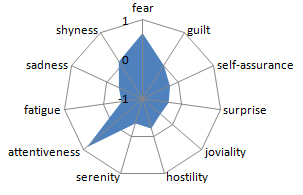}\label{fig:h1n1}}
			 \subfigure[AirFrance]{\includegraphics[width=0.32\textwidth]{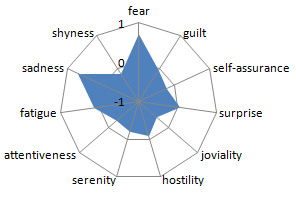}\label{fig:airfrance}}
			 \subfigure[US-Elec]{\includegraphics[width=0.32\textwidth]{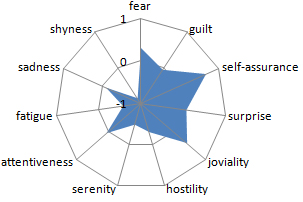}\label{fig:elect}}
			\subfigure[Obama]{\includegraphics[width=0.32\textwidth]{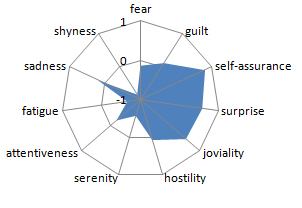}\label{fig:obama}}
			 \subfigure[MJ-death]{\includegraphics[width=0.32\textwidth]{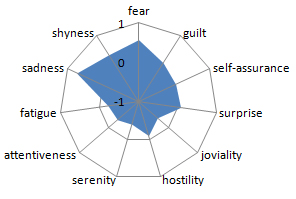}\label{fig:michaeljackson}}
			\subfigure[Susan-Boyle]{\includegraphics[width=0.32\textwidth]{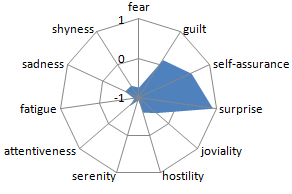}\label{fig:susanboyle}}
			\subfigure[Harry-Potter]{\includegraphics[width=0.32\textwidth]{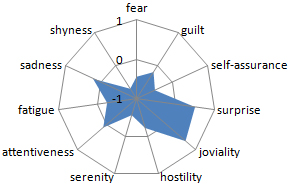}\label{fig:harrypotter}}
			 \subfigure[Olympics-begin]{\includegraphics[width=0.32\textwidth]{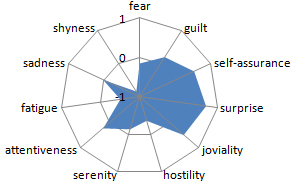}\label{fig:olympicsbegin}}
			 \subfigure[Olympics-end]{\includegraphics[width=0.32\textwidth]{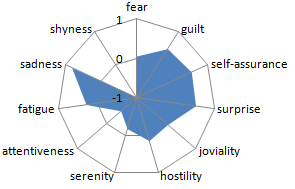}\label{fig:olympicsend}}
	\tiny{\caption{Events and feelings associated with them using PANAS-t.}}
	\label{fig:subFigures1}
\end{figure*}

In order to test how accurately PANAS-t can measure sentiment fluctuations, we calculated the PANAS-t scales for all events and present them in Kiviat representations. In each Kiviat
graph,  radial lines starting at the central point -1 represents each sentiment with the maximum value of 1~\cite{rajjain}. In Figure~\ref{fig:subFigures1}, we plot the eleven sentiments
in each figure so that each figure represents the corresponding event.

The first event we examine is \textsc{H1N1}, which represents the worldwide disease outbreak of the H1N1 influenza. The marking date, March 1st of 2009 was the day, where the influenza was declared by World Health Organization (WHO) as the global pandemic. To identify the event, we searched for a number of keywords including ``pandemic'' and ``swine'' and found a total of 335,969 relevant tweets during the five months period. Figure~\ref{fig:h1n1} shows the sentiment scores of this event based on PANAS-t scales. It demonstrates that the emotional state of Twitter users increased in \textit{attentiveness} (P(s) = 0.8774) and \textit{fear} (P(s) = 0.6768) in the days just after the announcement. Indeed, these two feelings correspond to the most likely feelings to expect from this event as people were both attentive to the precautions as well as afraid of a global pandemic.

The second event is \textsc{AirFrance}, which describes the tragic crash of an airplane on July 1st, 2009, which caused a big commotion in Twitter. The AirFrance Flight 447 was a scheduled as commercial flight from Rio de
Janeiro to Paris, but crashed in Ocean and killed all the 216 passengers. As expected, the crash caused sad emotions towards those who died and also fear that a something similar might happen again. Figure~\ref{fig:airfrance} shows the Kiviat representation for this event. As expected, \textit{fear} (P(s) = 0.72914) and \textit{sadness} (P(s) = 0.6992) were the two most predominated feelings in the tweets associated to this event.

The third event is \textsc{US-Elec}, which describes the presidential election related tweets in the US. With the election, many voters might feel apprehensive and even excited about the power of choice that is given to them. Our results show sentiments on this direction.  Figure~\ref{fig:elect} shows that users had the feeling for
\textit{self-assurance} (P(s) = 0.6741), \textit{joviality} (P(s) = 0.4277) and \textit{fear} (P(s) = 0.3072) increased, when the election results came out.

The fourth event, \textsc{Obama}, describes the president Barack Obama's inauguration speech, which received wide attention in Twitter. As reported in reference~\cite{HCD}, the majority of Americans were
more confident in the improvement of the country after viewing President Barack Obama's inauguration speech. Our analysis of the mood of Twitter's users performed on the day of
Obama's speech shows a particularly large increase in  \textit{self-assurance's} (P(s) = 0.7980), followed by \textit{surprise} (P(s) = 0.5802), and \textit{joviality}
(\textit{P(s) = 0.5227}). But despite all the positive manifestation regarding the election of Obama, we can also see a positive, but not so high value for \textit{sadness} (P(s) =
0.1789), which might naturally represent tweets from Barack Obama's oppositors.  Figure~\ref{fig:obama} shows that the feelings measured with PANAS-t are agreement with the ones
reported in reference~\cite{HCD}.

The fifth Kiviat chart, \textsc{Michael-Jackson}, is about the death of singer Michael Jackson. According to DailyMail~\cite{DailyMail}, nine of the ten most popular topics in Twitter were dedicated to the event the day after his death. In Figure~\ref{fig:michaeljackson}, we can see an increase in \textit{sadness} (P(s) = 0.4055), fear (P(s) = 0.5676), \textit{shyness} (P(s) = 0.4055), \textit{guilt} (P(s) = 0.1616), and \textit{surprise} (P(s) = 0.0810). It is interesting to perceive that, in addition to the expected feelings associated with a sudden death like \textit{sadness} and \textit{fear}, we could see increase in \textit{guilt}. This may be explained by the fact that many speculated about who or what killed Michael Jackson and fans and critics blamed the high stress caused by paparazzi and media for the death of celebrity.  Therefore, some Twitter users felt \textit{guilt} for his death and expressed such feeling in their tweets.

The next event we analyze is \textsc{Susan-Boyle}, who's appearance as a contestant the TV show, Britain's Got Talent,  had an incredible repercussion in the media. Global interest was triggered by the contrast between her powerful voice singing {``I Dreamed a Dream"} from the musical Les Miserables  and her plain appearance on stage. The contrast of the audience's first impression of her, with the standing ovation she received during and after her performance, led to an immediate viral spread over the social networks and a huge attention of the global media.  Figure~\ref{fig:susanboyle} shows that the sentiments expressed in Twitter associated with Susan Boyle's first appearance are \textit{surprise} (P(s) =  0.9066), followed by \textit{self-assurance} (P(s) = 0.4751), and \textit{guilt} (P(s) = 0.1367). The high surprise factor could also explain why Susan Boyle's video went viral on the Internet. People also felt self-assured as it is encouraging to see a woman successfully facing an audience that is laughing at her.  Finally, guilt is also expected as the event is based on wrong prejudice based on appearance.

The seventh event we studied is \textsc{Harry-Potter}, which describes the release of the movie ``Harry Potter and the Half-Blood Prince".  Figure~\ref{fig:harrypotter} shows that the main feelings associated are \textit{joviality} (\textit{P(s) = 0.6355}),  \textit{surprise} (P(s) = 0.4926), and \textit{sadness} (\textit{P(s) = 0.2056}), which also is described by many other critics that say the movie will leave the audience ``pleased, amused, excited, scared, infuriated, delighted, sad, surprised, and thoughtful.''

The last two charts shown in the figure are related to the \textsc{Olympics} games that were held in the summer of 2008 in Beijing, China. For this event, we show two Kiviat charts: one drawn based on the beginning sentiments and the other based on the ending sentiments of the event.  Figure~\ref{fig:olympicsbegin} is based on sentiments from the day of opening ceremony on August 08, where people felt  \textit{surprise} (\textit{P(s) = 0.7024}), \textit{attentiveness} (\textit{P(s) = 0.4621}), and \textit{joviality} (\textit{P(s) = 0.3298}). However, in the end of the event, on August 24th, we can see that these feelings had a decrease, whereas \textit{sadness} increased from \textit{P(s)= 0.1222} in this day and to \textit{P(s) = 0.5245} in the next day, as we can see in Figure~\ref{fig:olympicsend}.

\subsection{Testing across different geographical regions}

\noindent
In order to evaluate whether PNAS-t can effectively capture the subtle sentiment differences across different geographical areas, we take the example of the popular \textsc{H1N1} event and examine how sentiments on the event fluctuate over time in two different regions: USA and Europe.

To give further context of the \textsc{H1N1} event, we start by describing its impact on society. The H1N1 influenza, or also known as the ``swine flu'' by the public, has killed as many as half a million people in 2009. The World Health Organization (WHO) declare it as the first global pandemic since the 1968 Hong Kong flu, which caused a large concern in the world population. Later, WHO launched several warnings and precautions that should be taken by governments and by public, taking the entire population to a state of world alert against the disease.

In this section we compare the fluctuations of the mood of users about H1N1 in two different locations. More specifically, we want to verify how USA and European Twitter users felt about the event and quantify differences in public mood according to geographic regions.

In examining the difference in sentiments across North America and Europe, we focus on only English tweets. Therefore sentiments in Europe are limited to those tweets residing from Europe written in English. To be consistent in language representativeness, we limited our focus to tweets residing from the following regions in Europe: Ireland, Kingdom of the Netherlands, Malta, and United Kingdom. To do this we used a database collected and used in reference~\cite{Kulshrestha}. In this paper, authors used an expressive database from Twitter to separate unique ids, that represents users, by location. 

\begin{figure}[t]
		\centering
			 \subfigure[H1N1-Europe]{\includegraphics[width=0.35\textwidth]{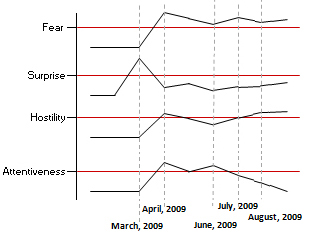}\label{fig:europe}}\\
			\subfigure[H1N1-US]{\includegraphics[width=0.35\textwidth]{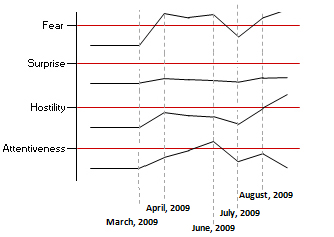}\label{fig:usa}}
	\caption{Public mood for H1N1 over 2009 in Europe and U.S.}
	\label{fig:subFigures2}
\end{figure}

The sparkline charts shown in Figures~\ref{fig:europe} and~\ref{fig:usa} present the fluctuations of four of the major sentiments related to the event in {PANAS-t} scales for Europe and USA, respectively. The charts are marked with five dates that indicate the day of important announcements made by WHO. In March, Mexican authorities begin picking up cases of what WHO called an ``influenza-like-illness." This event led European users to have an increase in the feeling of \textit{surprise} (\textit{P(s) = 0.8730}) but the same did not happen with users in the US.

In April, the first case of H1N1 in the United States was confirmed and WHO issued a health advisory on the outbreak of ``influenza like illness in the United States and Mexico'', and the charts shown a similar increase in \textit{fear} in both locations, \textit{P(s) = 0.7401} for Europe and \textit{P(s) = 0.6154} for the US.  We also see an increase in
\textit{attentiveness}, but this trend is only for Europe (\textit{P(s) = 0.4423}).

In June, WHO declared the new strain of swine-origin H1N1 as a pandemic, causing an increase of \textit{fear} (\textit{P(s) = 0.5385}) but also in \textit{attentiveness} in the US (\textit{P(s) = 0.3491}) and in users from Europe (\textit{P(s) = 0.3174}). In July, 26,089 new cases of H1N1 were confirmed in Europe by WHO, which leads to a further increase in sentiment of \textit{fear} (\textit{P(s) = 0.4887}), mainly among the European users.

On the last marked date in August, the most affected countries and deaths were announced as being located in Europe and America~\cite{who1}. In this period, European users had an increase in feeling of \textit{hostility} (\textit{P(s) = 0.2542}), whereas users in the US increased the feeling of \textit{fear} (\textit{P(s) = 0.4112}). These variations in the degree of sentiments expressed over time can effectively capture the dynamics in people's moods across different geographical regions.

\subsection{Testing across different time periods}

\begin{figure}[t]
		\centering
			 \subfigure[Samoa]{\includegraphics[width=0.35\textwidth]{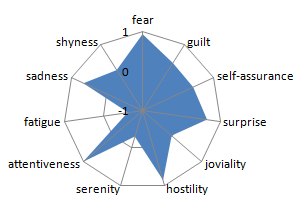}\label{fig:tsunami}}\\
			 \subfigure[Haiti]{\includegraphics[width=0.35\textwidth]{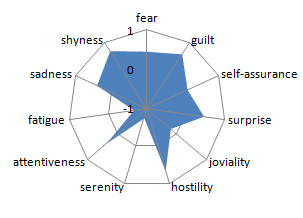}\label{fig:earthquake}}
	\caption{Feeling expressed by Twitter's users for Tsunami, in Samoa Islands, and Earthquake, in Haiti.}
	\label{fig:subFigures3}
\end{figure}

\noindent
The baseline values computed for {PANAS-t} in Table~\ref{tab:table3} is based on longitudinal data, based on 3.5 years worth of tweets between 2006 and until mid 2009, and represent a rather stable base sentiment of Twitter users. Therefore, these baseline values can be used to detect feelings of Twitter users from much later time periods (beyond mid 2009). Here, we use a different Twitter dataset that contains tweets posted between the end of 2009 to the end of 2010 that was collected by~\cite{Dredze} and have extracted tweets associated with two last events in Table~\ref{tab:table4}: \textsc{Samoa} and \textsc{Haiti}.

The 2009 \textsc{Samoa} Islands Tsunami was caused by a submarine earthquake that took place in the Samoan Islands on  September 29th with a magnitude of 8.1, which was the largest earthquake of 2009. A tsunami was generated causing substantial damage and loss of life in Samoa, American Samoa, and Tonga. More than 189 people were killed including children, which caused a large commotion around the world and generated a state of alert in neighboring coastal countries~\cite{digitaljournal}. Figure~\ref{fig:tsunami} shows the Kiviat chart for mood of users on the day of tsunami and the day after, which shows dominance in feelings of \textit{fear} (\textit{P(s) = 0.9280}), \textit{attentiveness} (\textit{P(s) = 0.9932}), \textit{hostility} (\textit{P(s) = 0.8451}), \textit{surprise} (\textit{P(s) = 0.6528}), and \textit{sadness} (\textit{P(s) = 0.6483}).

A similar tragic event happened in three months later in another part of the world. The 2010 \textsc{Haiti} earthquake was a catastrophic natural disaster, which caused severe damage in Port-au-Prince and the nearby region killing at least 250,000 people. Figure~\ref{fig:earthquake} shows that feelings of \textit{hostility} (\textit{P(s) = 0.9280}), \textit{attentiveness} (\textit{P(s) = 0.3678}), \textit{surprise} (\textit{P(s) = 0.4576}) and \textit{sadness} (\textit{P(s) = 0.3975}) had an increase. We also see an increase in \textit{shyness} and \textit{guilt}. After this event the world's eyes were focused on the disaster and people around the world offered help to Haiti~\cite{haiti}. As the poverty and precarious situation of the Haiti people was unveiled in the news, it is possible that this situation has generated an increase of these two feelings among the Twitter users. This finding demonstrates that PNAS-t is stable and can effectively represent sentiments of tweets gathered much later in time.

\section{Conclusions}

\noindent
In this paper, we present PANAS-t an eleven-sentiment psychometric scale adapted to the context of Twitter. PANAS-t is based on the expanded version of the well known Positive
Affect Negative Affect Scale (PANAS-x). Using empirical data from a unique Twitter dataset containing 1.8 billion tweets, we were able to compute the normalization scores for each
sentiment. We conducted a three-step evaluation. We first applied PANAS-t to 11 notable events that were widely discussed in Twitter. We next compared PANAS-t with a method using
most common emoticons that are used for users in Web. We finally showed that our method can be used in other database and also in other periods. These results provide strong
evidences that PANAS-t can accurately capture the positive and negative sentiments about events in Twitter.

The normalized scores of sentiments provided in this paper allow anyone to easily use PANAS-t, making it very simple and practical to be used
for large amounts of data and even for real-time analysis.  We hope that this psychometric scale can be used by any researches with the purpose of create tools that can be
used for government agencies or companies that might be interested in improving their products using social networks.  From the researcher perspective our method would allow one to
comprehend how, when, and why individuals feel and their feelings fluctuate according to social and economic events.

Despite the new opportunities our work brings, there are several limitations. First, the tweets we examined do no represent everyone who expressed sentiments in Twitter. We only focused on those tweets that explicitly contained ``I am feeling'' kinds of tags, although other tweets may contain emotions as well. Nonetheless, classifying emotional content from informational content remains an important challenge in social media analysis. 

Second, one criticism of sentiment analysis is that it takes a naive view of emotional states, assuming that personal moods can simply be divined from word selection. This might seem particularly perilous on a medium like Twitter, where sarcasm and other playful uses of language may subvert the surface meaning of a tweet. Deeper linguistic analysis should be explored to provide ``a richer and a more nuanced view'' of how people present themselves to the world.

We expect that in the future more applications will utilize sentiment analysis for specific vocabularies especially in a dynamic environment like Twitter to understand people's moods. Thus, we plan to combine other techniques such as machine learning to dynamically incorporate sentiments to PANAS-t according to the context.

\bibliographystyle{elsart-num-sort}
\bibliography{references}

\begin{thebibliography}{10}
\expandafter\ifx\csname url\endcsname\relax
  \def\url#1{\texttt{#1}}\fi
\expandafter\ifx\csname urlprefix\endcsname\relax\def\urlprefix{URL }\fi

\bibitem{Airoldi+Bai+Padman:06a}
E.~M. Airoldi, X.~Bai, R.~Padman, Markov blankets and meta-heuristic search:
  {Sentiment} extraction from unstructured text, Lecture Notes in Computer
  Science 3932 (Advances in Web Mining and Web Usage Analysis) (2006) 167--187.

\bibitem{DBLP:journals/corr/abs-1003-5699}
S.~Asur, B.~A. Huberman, Predicting the future with social media, CoRR
  abs/1003.5699.

\bibitem{Aue+Gamon:05a}
A.~Aue, M.~Gamon, Customizing sentiment classifiers to new domains: A case
  study, in: Proceedings of Recent Advances in Natural Language Processing
  (RANLP), 2005.

\bibitem{DailyMail}
C.~Bates, How michael jackson's death shut down twitter, brought chaos to
  google...and 'killed off' jeff goldblum, \url{http://bit.ly/16e6eM}, accessed
  January, 2012.

\bibitem{benevenuto@ceas10}
F.~Benevenuto, G.~Magno, T.~Rodrigues, V.~Almeida, Detecting spammers on
  twitter, in: Proceedings of the Annual Collaboration, Electronic messaging,
  Anti-Abuse and Spam Conference (CEAS), 2010.

\bibitem{Bermingham:2010:CSM:1871437.1871741}
A.~Bermingham, A.~F. Smeaton, Classifying sentiment in microblogs: is brevity
  an advantage?, in: Proceedings of the 19th ACM international conference on
  Information and knowledge management, 2010.

\bibitem{Bollen}
J.~Bollen, A.~Pepe, H.~Mao, Modeling public mood and emotion: Twitter sentiment
  and socio-economic phenomena, in: Proceedings of the International AAAI
  Conference on Weblogs and Social Media (ICWSM), 2011.

\bibitem{cha_icwsm10}
M.~Cha, H.~Haddadi, F.~Benevenuto, K.~P. Gummadi, {Measuring User Influence in
  Twitter: The Million Follower Fallacy}, in: Proceedings of the International
  AAAI Conference on Weblogs and Social Media (ICWSM), 2010.

\bibitem{Cheong}
M.~Cheong, V.~C. Lee, A microblogging-based approach to terrorism informatics:
  Exploration and chronicling civilian sentiment and response to terrorism
  events via twitter, Information Systems Frontiers 13 (2011) 45--59.

\bibitem{Chesley+al:06a}
P.~Chesley, B.~Vincent, L.~Xu, R.~Srihari, Using verbs and adjectives to
  automatically classify blog sentiment, in: AAAI Symposium on Computational
  Approaches to Analysing Weblogs (AAAI-CAAW), 2006.

\bibitem{benevenuto@ceas11}
S.~Chhabra, A.~Aggarwal, F.~Benevenuto, P.~Kumaraguru, Phi.sh/\$ocial: The
  phishing landscape through short urls, in: Proceedings of the 8th Annual
  Collaboration, Electronic messaging, Anti-Abuse and Spam Conference (CEAS),
  2011.

\bibitem{Diakopoulos:2010:CDP:1753326.1753504}
N.~A. Diakopoulos, D.~A. Shamma, Characterizing debate performance via
  aggregated twitter sentiment, in: Proceedings of the 28th international
  conference on Human factors in computing systems, 2010.

\bibitem{Dodds-Happiness}
P.~Dodds, C.~Danforth, Measuring the happiness of large-scale written
  expression: Songs, blogs, and presidents, Journal of Happiness Studies 11
  (2010) 441--456.

\bibitem{Golder.science.2011}
S.~A. Golder, M.~W. Macy, Diurnal and seasonal mood vary with work, sleep, and
  daylength across diverse cultures, Science 333~(6051) (2011) 1878--1881.

\bibitem{gomide2010dengue}
J.~Gomide, A.~Veloso, W.~M. Jr., V.~Almeida, F.~Benevenuto, F.~Ferraz,
  M.~Teixeira, Dengue surveillance based on a computational model of
  spatio-temporal locality of twitter, in: Proceedings of the ACM Web Science
  Conference (WebSci), 2011.

\bibitem{pcalaisKDD11}
P.~H.~C. Guerra, A.~Veloso, W.~Meira, Jr, V.~Almeida, From bias to opinion: a
  transfer-learning approach to real-time sentiment analysis, in: Proceedings
  of the ACM SIGKDD Conference on Knowledge Discovery and Data Mining (KDD),
  2011.

\bibitem{HCD}
HCD, Confidence levels increase among democrats and independents, decrease
  among republicans after viewing obama's press conference,
  \url{http://www.hcdi.net/news/MediacurvesRelease.cfm?M=276}, accessed January
  15, 2012.

\bibitem{digitaljournal}
Fear for new tsunami after earthquake hits sumatra,
  \url{http://digitaljournal.com/article/279868}.

\bibitem{who1}
Map of affected countries and deaths as of 23 august 2009,
  \url{http://www.who.int/csr/don/2009_08_28/en/index.html}.

\bibitem{haiti}
Operation compassion responds to haitian earthquake victims,
  \url{http://www.operationcompassion.org/2010/01/operation-compassion-responds-to-haitian-earthquake-victims/}.

\bibitem{iran.twitter}
{US confirms it asked Twitter to stay open to help Iran protesters},
  \url{http://tinyurl.com/klv36p}.

\bibitem{rajjain}
R.~Jain, The Art of Computer Systems Performance Analysis: Techniques for
  Experimental Design, Measurement, Simulation, and Modeling, 1st ed., John
  Wiley and Sons, INC, 1991.

\bibitem{Kim}
B.~E. Kim, S.~Gilbert, Detecting sadness in 140 characters: Sentiment analysis
  and mourning michael jackson on twitter, Web Ecology 03 (2009) 1--15.

\bibitem{Kulshrestha}
J.~Kulshrestha, F.~Kooti, A.~Nikravesh, K.~P. Gummadi, Geographic dissection of
  the twitter network, in: Proceedings of the International AAAI Conference on
  Weblogs and Social Media (ICWSM), 2012.

\bibitem{Lazer2009}
D.~Lazer, A.~Pentland, L.~Adamic, S.~Aral, A.~l\'{a}szl\'{o} Barab\'{a}si,
  D.~Brewer, N.~Christakis, N.~Contractor, J.~Fowler, M.~Gutmann, T.~Jebara,
  G.~King, M.~Macy, D.~Roy, M.~V. Alstyne, Computational social science,
  Science 323~(5915) (2009) 721--723.

\bibitem{Dredze}
P.~J. Michael, M.~Dredze, You are what you tweet: Analyzing twitter for public
  health, in: Proceedings of the International AAAI Conference on Weblogs and
  Social Media (ICWSM), 2011.

\bibitem{miyoshi:sentiment}
T.~Miyoshi, Y.~Nakagami, Sentiment classification of customer reviews on
  electric products, International Symposium in Information Technology (ITSim)
  (2007) 2028--2033.

\bibitem{citeulike:7044833}
B.~O'Connor, R.~Balasubramanyan, B.~R. Routledge, N.~A. Smith, From tweets to
  polls: Linking text sentiment to public opinion time series, in: Proceedings
  of the International AAAI Conference on Weblogs and Social Media (ICWSM),
  2010.

\bibitem{Pak}
A.~Pak, P.~Paroubek, Twitter as a corpus for sentiment analysis and opinion
  mining, in: N.~C.~C. Chair), K.~Choukri, B.~Maegaard, J.~Mariani, J.~Odijk,
  S.~Piperidis, M.~Rosner, D.~Tapias (eds.), Proceedings of the Seventh
  International Conference on Language Resources and Evaluation (LREC),
  European Language Resources Association (ELRA), 2010.

\bibitem{Pang+Lee:08b}
B.~Pang, L.~Lee, Opinion mining and sentiment analysis, Foundations and Trends
  in Information Retrieval 2 (2008) 1--135.

\bibitem{Pennebaker.liwc.2003}
J.~W. Pennebaker, M.~R. Mehl, K.~G. Niederhoffer, Psychological aspects of
  natural language use: Our words, ourselves, Annual Review of Psychology 54
  (2003) 547--577.

\bibitem{Sakaki@www10}
T.~Sakaki, M.~Okazaki, Y.~Matsuo, Earthquake shakes twitter users: real-time
  event detection by social sensors, in: Proceedings of the International
  Conference on World Wide Web (WWW), 2010.

\bibitem{Tumasjan}
A.~Tumasjan, T.~O. Sprenger, P.~G. Sandner, I.~M. Welpe, Predicting elections
  with twitter: What 140 characters reveal about political sentiment, in:
  Proceedings of the International AAAI Conference on Weblogs and Social Media
  (ICWSM), 2010.

\bibitem{Watson-X}
D.~Watson, L.~A. Clark, {The PANAS-X: Manual for the positive and negative
  affect schedule-Expanded Form}, {University of Iowa, 1994}.

\bibitem{Watson}
D.~Watson, L.~A. Clark, A.~Tellegen, {Development and validation of brief
  measures of positive and negative affect: the PANAS scales}, Journal of
  Personality and Social Psychology 54 (1988) 1063--1070.

\bibitem{twitter}
K.~Wickre, Celebrating twitter7,
  \url{http://blog.twitter.com/2013/03/celebrating-twitter7.html}, accessed on
  May 16, 2013.

\end{thebibliography}

\IEEEpeerreviewmaketitle

\end{document}